\documentclass[a4paper,11pt]{article}
\usepackage{pos}

\title{Chiral dynamics: Quo vadis~?}

\author*[a,b,c]{Ulf-G. Mei{\ss}ner}

\affiliation[a]{Helmholtz-Institut f\"ur Strahlen- und Kernphysik and Bethe
        Center for Theoretical Physics,
        Universit\"at Bonn,  D-53115 Bonn, Germany} 

\affiliation[b]{Institute for Advanced Simulation (IAS-4),  Forschungszentrum J\"{u}lich,
        D-52425 J\"{u}lich, Germany}

\affiliation[c]{Peng Huanwu Collaborative Center for Research and Education, Beihang University, Beijing 100191, China}

\emailAdd{meissner@hiskp.uni-bonn.de}

\abstract{I review the status of chiral dynamics. Topics include pion-pion scattering, dynamically
generated states in  the hadron spectrum and the emergence of two-pole structures, chiral symmetry 
in nuclear physics and chiral dynamics in the Big Bang.}

\FullConference{The 11th International Workshop on Chiral Dynamics (CD2024)\\
 26-30 August 2024\\
Ruhr University Bochum, Germany\\}

\begin{document}
\maketitle

\section{Introductory remarks}

I start this talk with an anecdote. To my knowledge, one of the first papers that included
``chiral perturbation theory''  (in a slightly modified version)  in the title was Ref.~\cite{Honerkamp:1971xtx} 
from Bonn, which states in the first paragraph: ``Although we do not believe that it is likely to
be useful from a  physical point of view to pursue the perturbation theory of chiral-invariant
Lagrangians, ...''. This seems to make this talk obsolete. Of course, that paper was written before the
emergence of effective field theories, which gave a very different meaning to the issue
of renormalization, which was the main topic of Ref.~\cite{Honerkamp:1971xtx}. In contrast to what was
believed at that time, we know now that all fundamental quantum field theories are indeed effective field theories.

This brings me to the
next basic issue, namely what precisely does chiral dynamics  mean? This wording was
first used by Julian Schwinger in Ref.~\cite{Schwinger:1967tc}, who replaced the cumbersome current
algebra operator techniques by a numerical effective Lagrange function, which ultimately led to his source theory.
I recommend the introduction to Weinberg's contribution to the  symposium honoring Julian Schwinger on the 
occasion of his 60th birthday for a historical perspective of this work~\cite{Weinberg:1978kz}.
The  terminology ``chiral dynamics'' is frequently used since then, but often with a  different meaning.
The wording ``chiral perturbation theory'' was used first in this strict
form by Langacker and Pagels~\cite{Langacker:1973hh} but only became a household issue
due to Gasser and Leutwyler~\cite{Gasser:1983yg,Gasser:1984gg}. Thus, in what follows,
I will use the following definitions:
\begin{itemize}
 \item {\bf Chiral perturbation theory (CHPT)} refers to a
strict perturbative expansion in (a) small parameter(s), like the light quark masses and/or small 
momenta/energies.
\item {\bf Chiral dynamics (CD)} is used when  some non-perturbative resummation is involved,
say the kernel of a scattering process can be expanded using the CHPT counting rules but then
the scattering matrix is solved non-perturbatively.
\end{itemize}

So in this talk, I will address issues in chiral dynamics. Of course, other persons
use other definitions, thus it is important to clearly spell out what one is
talking about.  In any case, CHPT and CD explore the spontaneously and explicitly broken chiral
symmetry of QCD and thus deepen our understanding of the Standard Model at low energies.
In this talk, I review a number of recent developments. This is clearly based on a subjective
choice, and other developments such as CHPT with axions, see the contribution by Feng-Kun Guo~\cite{talkFKG},
are covered in many other interesting talks at this workshop.  

\section{Pion-pion scattering: The poster child and its offsprings}

Pion-pion scattering (in the threshold region) is the purest process in two-flavor CHPT (and also
chiral dynamics) because the up and down quarks are really light on the
scale $\Lambda_{\rm QCD}$. I had talked about this fundamental reaction
already at CD2012~\cite{Meissner:2012ku} and at CD2018~\cite{Meissner:2019bsu},
so I can be short on the basics and mostly (but not only) discuss progress in lattice QCD.

At threshold, the $\pi\pi$ scattering amplitude is given in terms of two numbers,
the S-wave scattering lengths $a_0$ and $a_2$, corresponding to  total isospin zero
and two, respectively. The CHPT predictions  are (I concentrate here on $a_0$):
$a_0=0.16$~(LO)~\cite{Weinberg:1966kf},
$a_0 = 0.20\pm 0.01$~(NLO)~\cite{Gasser:1983yg}
$a_0 = 0.217\pm 0.009$~(NNLO)~\cite{Bijnens:1995yn}\footnote{Note that in that paper, no error was given but
two different solutions. I use here solution~1 as the central value and the difference to solution~2 as the uncertainty.}. 
The fairly large corrections at NLO are understood from the  strong pionic final-state interactions (FSI)
in this channel  and there are still
sizeable corrections at NNLO despite the small expansion parameter $(M_\pi/\Lambda_\chi)^2
\simeq 0.02$ for $\Lambda_\chi = 1\,$GeV.

The first offspring to improve on these
works was the combination of CHPT with dispersion relations, which involved
quite a number of  people and works that I possibly can not discuss here. 
Matching the 2-loop CHPT  representation of the $\pi\pi$ scattering amplitude  to the Roy equation 
solution allows to make the prediction for $a_0$  much more precise, namely
$a_0 =  0.220 \pm 0.005$~\cite{Colangelo:2000jc}. Even more, the dispersive analysis
allowed also to precisely determine the mass and width of the lowest resonance
in QCD, $M_\sigma = 441^{+16}_{-8}\,$MeV and $\Gamma_\sigma/2 =  272^{+9}_{-13}\,$MeV~\cite{Caprini:2005zr},
nowadays called $f_0(500)$.

A second offspring was the development of NREFTs to give a different
access to low-energy $\pi\pi$ scattering, more precisely, scattering at zero energy.
This led to a precision theory for hadronic atoms, 
which is relevant for  the $\pi\pi$, $\pi K$, $\pi N$, $\pi d$, $Kp$, $Kd$ systems in the
sector of the light quarks $u,d,s$.
For a review, see Ref.~\cite{Gasser:2007zt}. For the case of pionium (the $\pi^+\pi^-$ bound
state) experiment gives the scattering length combination $|a_0-a_2| =
0.2533^{+0.0107}_{-0.0137}$~\cite{Adeva:2011tc}, which is consistent
with the predictions~\cite{Colangelo:2000jc}. I mention in passing that the experimental
result of the the $\pi K$ atom, $ |a^{1/2}_0  - a^{3/2}_0|/3
= 0.072^{+0.031}_{-0.020}$~\cite{Yazkov:2018oel} is incidentally consistent with the one-loop
CHPT result  $0.073(2)$~\cite{Bernard:1990kw}, but space does not allow for a more in-depth
discussion on this fundamental reaction in three-flavor CHPT.  For recent work on $\pi K$ scattering
with many references, see e.g.~\cite{Pelaez:2020gnd}.

Another  offspring is the analysis of the cusp that in $K\to 3\pi$ that can
also be used to extract the pion-pion scattering lengths. Combining these 
with the FSI in $K_{e4}$ decays leads to
$a_0 = 0.2210\pm 0.0047_{\rm stat}\pm 0.0040_{\rm sys}$~\cite{NA482:2010dug}, again
in nice agreement with theory.

Thus, experiment and chiral dynamics are well aligned,
but what about lattice QCD (LQCD)? Recall that at the time of CD2012, no direct lattice $a_0$ determinations
were available (due to the difficulty in taming the disconnected diagrams) and at the time of
CD2018, two unquenched QCD simulations at unphysical pion masses were reported but the
errors appeared too small. Now a number of better simulations are available, however,
chiral extrapolation are still  needed in most cases. The recent state of the art in the 
lattice determinations of the pion-pion S-wave scattering lengths
with summary  figures/tables  can be found in Refs.~\cite{Bruno:2023pde,RBC:2023xqv}.
I focus here on the work of the GWU group~\cite{Mai:2019pqr}
as their work concerns the S- and P-waves including the pertinent resonances
(for a review on resonances from lattice QCD, see~\cite{Mai:2022eur}). They find
for the  $I=0$ S-wave $a_0 = 0.2132^{+0.008}_{-0.009}$ and a complex $f_0(500)$ mass of
$M_\sigma = (443(3)-i\,221(6))\,$MeV. While the central value of the mass is fine and in good
agreement with the dispersive analysis whereas the width is somewhat small, I consider the errors underestimated. 
That can also be seen from the too small  $\rho$ mass, $M_\rho = (724^{+2}_{-4} - i\, 67^{+1}_{-1})\,$MeV,
where the real part is many $\sigma$ away from the empirical value Re\,$M_\rho = 761 - 765$~MeV
\cite{ParticleDataGroup:2024cfk}. Note further that a recent calculation at the physical point
gives $M_\rho = (796(5)(50) -i\, 96(5)(16))\,$MeV, which agrees with the PDG values within the
sizeable errors~\cite{Boyle:2024hvv}, but has very different central values. So it is not
yet time to declare victory, more
LQCD work is certainly needed to precisely pin down even the lowest resonances in QCD.

\section{Novel insights into the hadron spectrum from chiral dynamics}

In  CHPT, resonances are a limit, not active degrees of freedom. In very few cases, one can extend the chiral effective Lagrangian to include resonance fields explicitly (see below), whereas CD allows to investigate a larger number of
unstable hadrons. In any case, one has to be aware of the decoupling theorem: The leading non-analytic terms stem 
from Goldstone boson one-loop graphs coupled to Goldstone bosons or ground state baryons~\cite{Gasser:1979hf}.
This has a number of consequences. First, resonances must decouple, which is sometimes not accounted for.
Second, the cuts and poles generated by the Goldstone bosons  (GBs) are not affected by resonances. Still, the resonances
leave their traces in saturating most of the low-energy constants (LECs) and further, the QCD chiral and large-$N_C$ limits do not commute. Note that in the large-$N_C$ limit, the nucleon and the $\Delta(1232)$ are degenerate in mass and thus the decoupling 
theorem is explicitly violated.

There  are essentially three ways of including resonances: First, the inclusion as {\em matter fields} is possible in a few cases, here one mostly  deals with the $\Delta(1232)$, see my talk at CD2018. This clearly requires an
extended power counting and/or the complex-mass scheme. Second, in some cases, {\em CHPT combined with
dispersion relations} (DRs)  allows to study resonances, such as the already mentioned $f_0(500)$ as well as the $f_0(980)$, $\rho(770)$, $K_0^*(700)$, $K^*(890)$ in the meson sector and the $\Delta(1232)$ and the Roper $N^*(1440)$ in the baryon sector. Of course, DRs are a fine tool to study resonances in general, but this is mostly  limited by the available data on the pertinent scattering processes, as witnessed by the renowned work of the Karlsruhe-Helsinki group many decades ago~\cite{Hohler:1979yr}. That work set a standard on the extraction of resonance properties which is unfortunately 
not always  achieved  in present day analyses. Here, I will give one example of recent work in this field and refer for more details to the talk by Jacobo Ruiz de Elvira in these  proceedings~\cite{talkJRE}. Third,  single channel unitarization or the more general   {\em coupled-channel chiral dynamics} (non-perturbative unitarity)
allows to study certain resonances ($\rho$, $\sigma$,..) and, in particular, to deal with the strange 
baryons~\cite{Kaiser:1995eg}. For a nice review with a historical perspective, see~\cite{Oller:2020guq}.

The Roy equation program for $\pi\pi$ scattering  can also be performed in the pion-nucleon system, where the 
basic equations are the so-called Roy-Steiner equations, see Ref.~\cite{Hoferichter:2015hva} and references therein.
First, this allows for a high-precision determination of the $\pi N$ $\sigma$-term including isospin-breaking
corrections, $\sigma_{\pi N} =59.0(3.5)\,$MeV~\cite{Hoferichter:2023ptl}. Second, the dimension-two
$\pi N$ LECs $c_{1,2,3,4}$ can be determined precisely, namely $c_1=1.10(3)$, $c_2=3.57(4)$, $c_3=-5.54(6)$
and $c_4 = 4.17(4)$, all in GeV$^{-1}$, and similarly for some of the dimension-three LECs $\bar{d}_i$, namely
$\bar{d}_1+\bar{d}_2$, $\bar{d}_3$, $\bar d_5$ and $\bar d_{14}-\bar d_{15}$.
These will be needed in the discussion of chiral symmetry in nuclei, see below. 
Third, the lowest nucleon resonances and their  couplings to various mesons ($f_0(500), f_0(980)$, and $\rho(770)$)
can be extracted from the Roy-Steiner analysis, such as
$M_\Delta = (1209.5(1.1) + i\, 98.5(1.2))\,$MeV and $M_R = (1374(3)(4) + i\, 215(18)(8))\,$MeV. It is important to
stress that the couplings of the various mesons to these baryons are complex-valued, as is the case for any unstable
state. For more details see Ref.~\cite{Hoferichter:2023mgy}.

\begin{figure}[h!]
\centering\includegraphics*[height=1.45cm,angle=0]{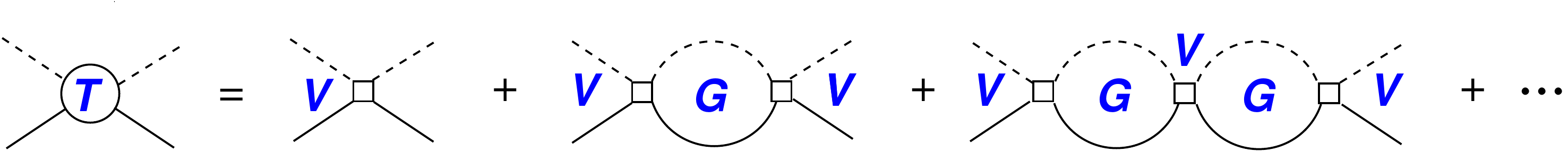}
\caption{Unitarization of GB scattering  (dashed lines) off baryons (solid lines) .}
\label{fig:bubble}
\end{figure}

The basic  idea of coupled-channel chiral dynamics can be spelled out easily. Consider a given $2\to 2$ scattering
process that involves a number of coupled channels.  In the first step, one uses CHPT\footnote{Or some combination
of CHPT and heavy quark symmetries, e.g. Goldstone bosons scattering off the D-meson triplet, or other types of extensions.} 
to construct the potential $V= V_{\rm LO} + V_{\rm NLO}+ ...$, which is then resummed, often in 
the on-shell approximation, $T = V /[1 +G V]$, where $T$ is the scattering matrix and $G$ the pertinent
two-hadron loop function, see Fig.~\ref{fig:bubble}. 
Note that $T$, $V$ and $G$ are matrices in channel space and any channel  index is suppressed for simplicity.
 The two-hadron loop function and the resummation procedure require regularization, and this
seems to introduce some model-dependence. It can, however,  be overcome by going to sufficiently high orders. 
More importantly, the resummation 
allows for the generation of resonances, in particular the elusive $\Lambda(1405)$, see Refs.~\cite{Mai:2020ltx,Hyodo:2020czb} for reviews. Before discussing this particular state, let me
point out that such dynamically  generated resonances are an integer part of the hadron spectrum, thus
any model, that does not allow for the inclusion of such states, is not a faithful representation of QCD and
thus should be abandoned. In particular, the analysis fo $\bar KN$ scattering in unitarized CHPT has revealed
a new class of players in the hadron spectrum, the so-called two-pole structures. See the talk by 
Lisheng Geng in these proceedings~\cite{talkLSG}. The terminus ``two-pole structure''  refers to the fact that
 particular single states in the hadron spectrum as listed in the PDG tables are indeed two states. In terms
 of unitarized CHPT, this was first observed in Ref.~\cite{Oller:2000fj} in a re-analysis of coupled-channel $K^- p$ scattering.
In that work, a number of technical improvements were introduced, such as the subtracted meson-baryon loop function
 with dimensional  regularization, which has become a standard methodology by now, the coupled-channel approach to the 
 $\pi\Sigma$ mass distribution and matching formulas to any order in chiral perturbation theory were established. Most importantly, it was found 
 that the $\Lambda(1405)$ is indeed described by two poles with rather different imaginary parts, that exhibit a clear departure from the Breit-Wigner (BW) situation (the latter issue is of particular importance for experimental collaborations that often misuse the  BW parameterization). The emergence of these two poles starting from the SU(3) limit of three-flavor QCD was 
 worked out in~\cite{Jido:2003cb}. When all quark masses are equal, $m_u=m_d=m_s$, the masses of the mesons
 in the GB octet are all the same and all baryon masses in the ground state octet are equal, thus all the baryons
are stable and thus reside on the real axis.  Consider the scattering of the GBs off the octet baryons.
Simple group theory gives $8 \otimes 8 = 1\oplus 8_s\oplus 8_a\oplus 10 \oplus  \overline{10} \oplus 27$, where one has binding at LO in the singlet and the two octets. This generates 17 mostly degenerate bound states (the two octets
are accidentally degenerate and the singlet is somewhat lighter). From these 17 bound states, only
 a few survive when one switches on the SU(3) symmetry breaking, in particular two states with $I=0$ in the vicinity
 of the $\Lambda(1405)$, one close to the $\bar KN$ and the other close  to the $\pi\Sigma$ threshold, respectively.  
 This is how CD generates these two poles (and a few others).  The two-pole structure of the   $\Lambda(1405)$
has been verified by many groups world-wide. However,  the lower pole is not yet precisely determined. In the PDG
tables, now two poles appear, the two-star resonance  $\Lambda(1380)$ (the low-mass pole) and the four-star resonance $\Lambda(1405)$ (the high-mass pole). Even more interesting,
 more of such two-pole structures have been found, as reviewed in Ref.~\cite{Meissner:2020khl}
and discussed here by Geng~\cite{talkLSG}. So the chiral dynamics  leaves clear imprints in the hadron spectrum,
which is certainly a very unexpected development and it is an open question how many of these two-pole structures will indeed be found.

\section{Chiral symmetry in nuclear interactions and  in nuclei}

\begin{figure}[h!]
\centering\includegraphics*[height=8.1cm,angle=270]{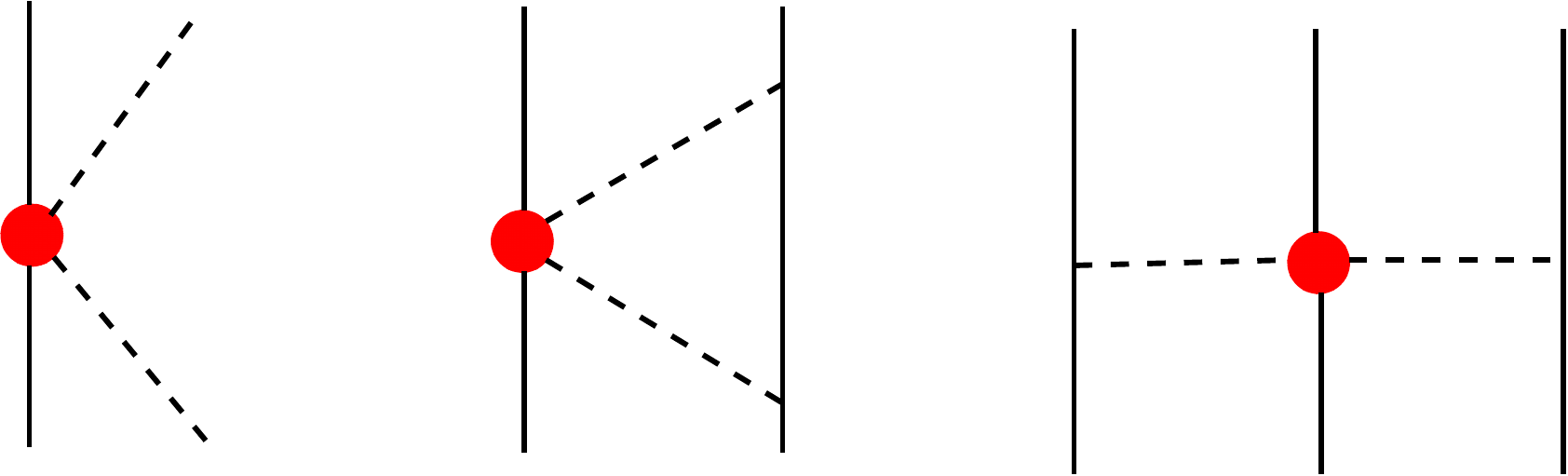}
\caption{The LECs $c_i$ (red circles) in pion-nucleon scattering (left), the two-pion exchange potential in the NN (middle)
and the 3N (right) forces, respectively. Solid (dashed) lines denote nucleons (pions).}
\label{fig:ci}
\end{figure}

As it is well-known, the pion was introduced by Yukawa in 1935 as the carrier of the strong force~\cite{Yukawa:1935xg}, and
was found in emulsion experiments about a decade later~\cite{Lattes:1947mx}. Only in the early 1970ties, the pion 
was firmly established
in nuclei by resolving the long-standing discrepancy between  the theoretical predictions for the threshold
neutron capture reaction, $n+p\to d+\gamma$,  in the impulse
approximation, $\sigma_{\rm IA} = 302.5(4.0)\,$mb, and the measured value $\sigma_{\rm exp} = 334.2(5)\,$mb,
as meson-exchange currents (MECs) just provide the missing 10\%~\cite{Riska:1972zz}. Such pionic  MECs
now appear naturally in the chiral EFT of nuclear forces and currents, see e.g. the early work in~\cite{Park:1993jf}.
Then came the 1993 shock, when Bertsch, Frankfurt and Strikman questioned the role of nuclear pions by
analyzing a number of medium- and high-energy experiments~\cite{Bertsch:1993vx}. Quickly, a number of 
(questionable?) solutions to recover the pions was published, but let me return back to chiral symmetry. As
pointed out so clearly by Weinberg (and others), chiral symmetry breaking in QCD relates {\em many} processes.
One of the best examples are  the dimension-two (three) vertices from the effective $\pi N$ Lagrangian 
$\sim c_i\,(\bar{d}_i)$ already
mentioned above. The corresponding  LECs can be precisely determined in pion-nucleon scattering and leave their traces in the two- as well as the three-nucleon interactions, see Fig.~\ref{fig:ci}.
Here, I concentrate on the NN interaction. The two-pion exchange (TPE) was already studied in the framework
of the chiral NN forces in \cite{Ordonez:1995rz,Kaiser:1998wa}, but a truely quantitative description was only
achieved later with the N4LO~\cite{Epelbaum:2014sza} and N4LO$^+$ potentials~\cite{Reinert:2017usi}. The 
leading two-pion exchanges $\sim c_i$ and $\sim d_i$ appear at N2LO and N4LO, leading to 
parameter-free contributions at N2LO and N4LO from TPE. In both cases, there is a clear improvement
in the description of the $np$ and $pp$ scattering data when going from NLO to N2LO and from N3LO to N4LO, respectively, due to this parameter-free TPE, see e.g. Table~III in Ref.~\cite{Reinert:2017usi}. Clearly, this is not an absolute measure as
the potential is not an observable.

Still, one might ask the question whether pions are really required in nuclear structure? First, many 
relativistic mean-field models based mostly on  $\sigma,\omega,\rho$ exchanges work rather well,
but they are not consistent with chiral symmetry and thus have no foundation in QCD. 
Furthermore, one can formulate nuclear physics based on EFTs just employing  contact interactions without any pions. Such  pionless EFT approaches are also not  constrained by 
chiral symmetry. I mention here the EFT build around the so-called unitary limit~\cite{Konig:2016utl} or the
so-called minimal nuclear action of NLEFT~\cite{Lu:2018bat} based on Wigner's SU(4) symmetry~\cite{Wigner:1936dx},
that allows one to describe neutron matter up to saturation density and the ground state properties of nuclei up to calcium with only four parameters. In this approach,
the spectrum of carbon can also be well described~\cite{Shen:2022bak} and the data on the $^4$He transition
form factor are reproduced precisely~\cite{Meissner:2023cvo}, see also the talk by Dean Lee~\cite{talkDL} in these
proceedings. So pions don't seem to be needed? They are, because there are  much different nuclear systems
that can not be described by these methods and also, the  precision is limited. To overcome the sign oscillations
that prevented NLEFT calculations beyond N2LO, which limits the precision of the calculations,
the new quantum many-body method of {\em wavefunction matching} was introduced in Ref.~\cite{Elhatisari:2022zrb},
see also~\cite{talkDL}.  In a nutshell, wavefunction matching (WFM) transforms the high-fidelity  interaction
(in our case the N3LO
chiral nuclear Hamiltonian)   between particles (in our case nucleons)  so that the wave functions of the high-fidelity Hamiltonian up to some finite range matches that of an easily computable Hamiltonian, where the latter gives an
approximate solution of the many-body problem and is largely free of the disturbing sign oscillations. More precisely, wavefunction matching operates entirely in the two-nucleon sector. For the nuclear case, this simplified Hamiltonian
consists of Wigner SU(4) symmetric two-nucleon forces and properly regularized one-pion exchange, and it is treated fully non-pertubatively. To bring the chiral Hamiltonian $H_\chi$ close to the simplified Hamiltonian $H_S$,
a unitary transformation is performed leading to $H_\chi' = U^\dagger H_\chi U$, and the differences to the full chiral Hamiltonian,  $H_\chi'-H_S^{}$, are then calculated in first order perturbation theory. Finally, fitting the various locally and non-locally smeared 3NF operators to the nuclear binding energies with $3\le A \le 58$, one can predict the corresponding nuclear charge radii as well as the equation of state of pure neutron as well as nuclear matter. All of these quantities agree with the data. This
would not be possible without the OPE in the simple Hamiltonian, so {\bf pions are indeed needed in nuclear structure}. The form of the simple Hamiltonian also gives further credit to Weinberg's power counting in chiral nuclear EFT.

 \begin{figure}[t!]
\centering\includegraphics*[height=6.5cm,angle=0]{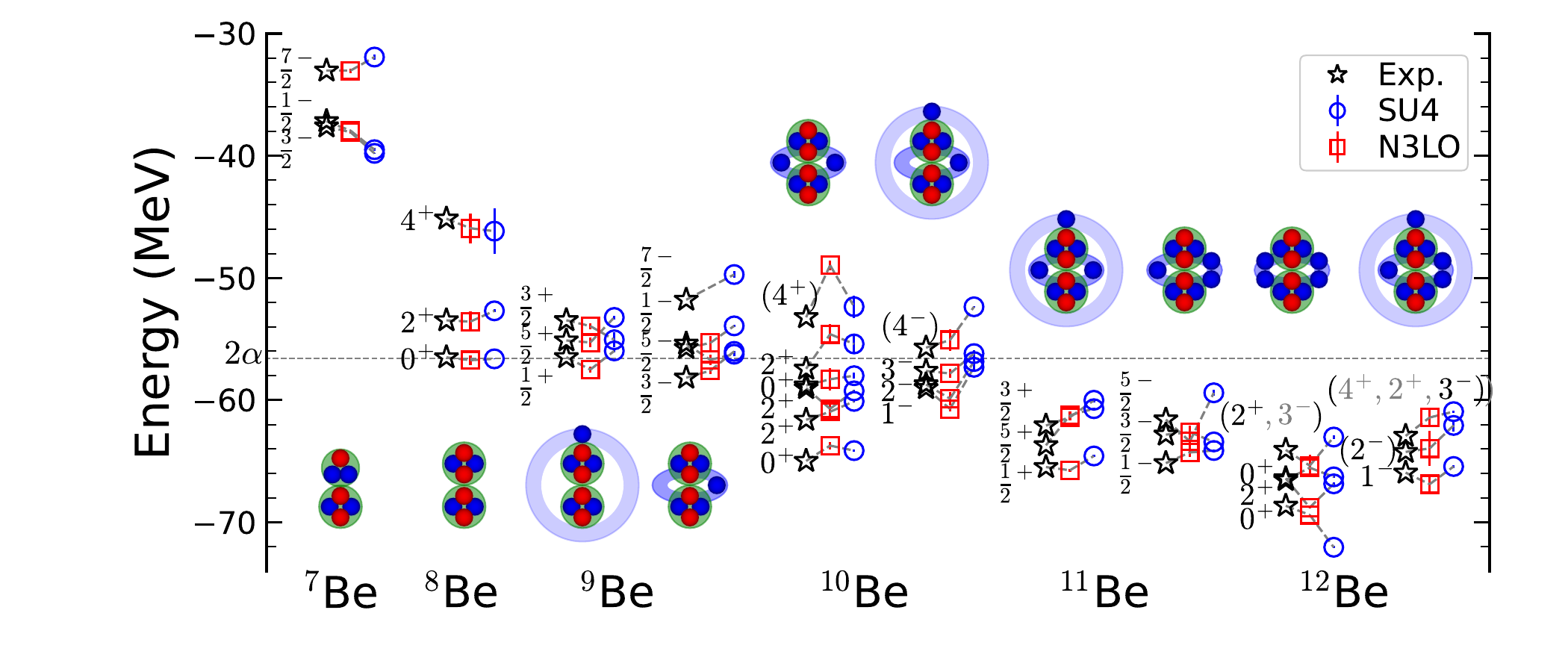}
\caption{Low-lying spectrum from $^7$Be to $^{12}$Be calculated by NLEFT
  using the N3LO interaction~\cite{Elhatisari:2022zrb} and  the SU(4) interaction~\cite{Shen:2022bak},
   compared to the data.
  The error bars correspond to one standard deviation errors include stochastic
  errors and uncertainties in the Euclidean time extrapolation.
  The two $\alpha$ threshold is denoted by horizontal dashed line.
  The cartoons display the dominant structure of each isotope. Figure courtesy of Shihang Shen.}
\label{fig:Be}
\end{figure}
As an application of the power of WFM, in Fig.~\ref{fig:Be} I display the low-lying states of the p-shell 
beryllium isotopes from $^{7}$Be to $^{12}$Be using the state-of-the-art full N3LO interaction and also 
the SU(4) interaction~\cite{Shen:2024qzi}. One finds  good agreement of the energies, radii, and electromagnetic 
properties with the data.  Clearly, the SU(4) approach does not precisely give these energies (as mentioned above), 
but still works astonishingly well.  More interestingly, the different geometrical properties of these isotopes come
out consistent with what is known, namely two-center cluster structures as well as one- and
two-neutron halos (as shown by the cartoons in Fig.~\ref{fig:Be}, but worked out in more detail using tomographic 
methods in~\cite{Shen:2024qzi}).  These results are quite intriguing because generally, different approaches
are used for these different structures, like cluster models or halo-EFT.

\section{Chiral dynamics in the Big Bang}

The light elements are generated in the Big Bang during the first 15 minutes of the Universe, which is called Big Bang Nucleosynthesis (BBN). This reaction network is characterized by a number of
fine-tunings, in particular let me mention the so-called deuterium bottleneck, where the produced deuterium is 
disintegrated by the abundant photons until the Universe has cooled down sufficiently.  This is by the way very 
different to high-energy heavy ion collisions, which are often described as the Big Bang in the laboratory.
But let me come back to the primordial nucleosynthesis.
These  can indeed be used to set limits on the possible variations of the fundamental constants of the 
Standard Model, in particular the light quark masses and the electromagnetic fine-structure
constant $\alpha$, an idea first entertained by Dirac~\cite{Dirac:1937ti}. For  recent reviews on these and
related issues, see Refs.~\cite{Adams:2019kby,Uzan:2024ded}.

Here, I report on some recent results obtained in collaboration with Bernard Metsch and
Helen Meyer, see also Helen Meyer's talk at this workshop~\cite{talkHM}. First, 
we obtained new results for  the dependence of the primordial nuclear abundances as a function of $\alpha$, 
keeping all other fundamental constants fixed~\cite{Meissner:2023voo}. This included  updates on  the 
leading nuclear reaction rates, and the inclusion of  the temperature-dependence of the leading nuclear reactions rates. Furthermore, the systematic uncertainties  were assessed by using five different publicly available codes for BBN. The 
current values for the observationally  based $^2$H and $^4$He abundances restrict the fractional change in $\alpha$ 
to less  than 2\%, which is a tighter  bound than found in earlier works on the subject.

\begin{figure}[h!]
\centering\includegraphics*[height=3.9cm,angle=0]{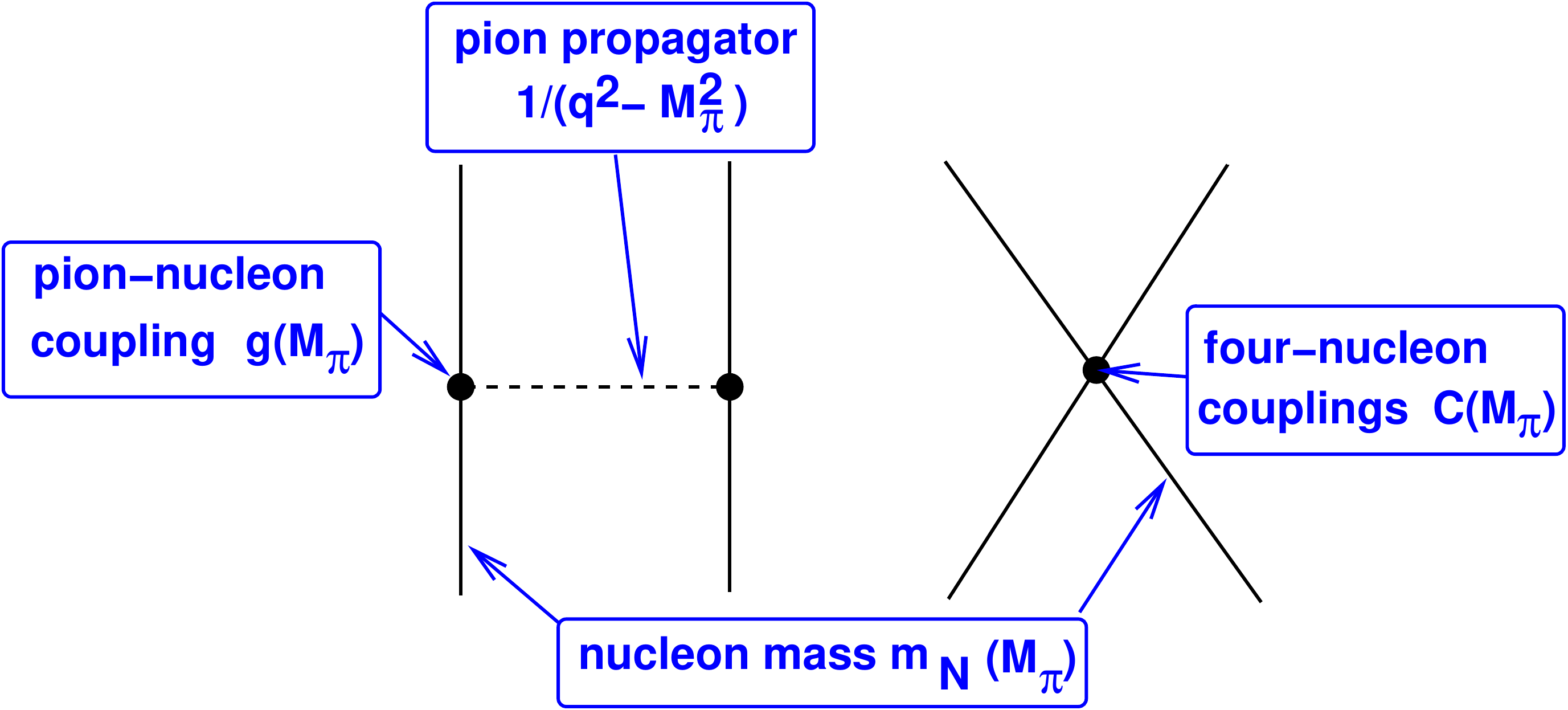}
\caption{Pion mass dependence of the NN interaction through the  OPE and the LO contact interactions. Solid (dashed) lines denote nucleons (pions).}
\label{fig:Mpi}
\end{figure}

More interesting from the view of chiral dynamics are the bounds on the quark mass variations, which 
has been a topic of many papers, see  e.g. Appendix~B in Ref.~\cite{Meyer:2024auq}.
In general, the pion mass dependence in nuclear systems can 
be most easily understood by looking at the LO nucleon-nucleon (NN) interaction in the Weinberg scheme, see Fig.~\ref{fig:Mpi}.
Due to the Gell-Mann--Oakes--Renner relation, $M_\pi^2 = B_0^{}(m_u^{}+m_d^{})$, 
the light quark mass dependence can be mapped onto the 
pion mass dependence, which is either explicit (pion propagator) or implicit (nucleon mass, pion-nucleon coupling, 4N LECs). In the most recent work~\cite{Meyer:2024auq},  we considered the possible
variations of the Higgs VEV $v$. Keeping, as mostly done, the Yukawa couplings fixed, the variation of the light
quark masses is proportional to the variation of  $v$. In this work, we included in particular the
new value of the  strong neutron-proton mass splitting from Ref.~\cite{Gasser:2020mzy}, $(m_n-m_p)_{\rm QCD} 
= (1.87\pm 0.16)\,$MeV,
and for the NN contact terms and deuteron binding energy we combined LQCD data with the low-energy 
theorems from Refs.~\cite{Baru:2015ira,Baru:2016evv}. Here, the detailed balance between the capture
$n+p\to d+\gamma$ and dissociation $\gamma +d\to n+p$  reactions is very sensitive to the change in the deuteron
binding energy, leading to very stringent bounds on possible variations of  the Higgs VEV, namely
$\delta v / v \in [ -0.0069, 0.0039]$ from the $^4$He abundance and $\delta v / v \in [-0.0007, -0.0002]$ from
the $^2$H abundance (by comparing to the PDG values). Two points are in particular noticeable:
In contrast to most earlier investigations, the deuterium abundance sets a stronger bound than the
one from $^4$He, and second, these bounds are much tighter than earlier ones, see e.g. the
recent work~\cite{Burns:2024ods}.

\section{Summary \& outlook}

Although CD is a mature field, still some basic predictions are just getting  precisely tested now
like the chiral anomaly in the process $\gamma \to 3\pi$, see the talk by Maltsev for the
COMPASS collaboration~\cite{talkAM}. As it is usually the case in this field, there is a strong
interplay between theory and experiment, see the talk by Bai-Long Hoid~\cite{talkBHL}.
There is also an on-going tension between the pion-nucleon $\sigma$-term from the RS
equations and lattice QCD determinations, see the talk by Jacobo Ruiz de Elvira~\cite{talkJRE},
which adds to the tensions discussed in the context of pion-pion scattering.  To my opinion, these 
can only be resolved by more lattice work, accounting for all possible systematic effects such
as excited state contaminations, see e.g.~\cite{Gupta:2021ahb}.

In the hadron spectrum, the emergence of two-pole structures is tightly connected to the
remaining ground state SU(3)$_V$  symmetry  of QCD and its  
explicit symmetry breaking, which is an integer part of chiral dynamics. This is
a new and fairly unexpected manifestation of chiral dynamics, which is based on a fruitful interplay of
experiment, theory and LQCD. The coupled-channel methods developed in this context 
can also serve to better analyze experimental as well as lattice data, see e.g. Ref.~\cite{Asokan:2022usm}.
Experimenters often use Breit-Wigner parameterizations in situations were they are not applicable. In fact, many 
resonance  properties in the PDG tables are obtained in that way, which requires modifications.

That pions play an important role in nuclei is  known since long~\cite{Ericson:1988gk}, in particular
through the excitation of the $\Delta(1232)$ resonance. There have been recent attempts
to include the  $\Delta(1232)$  in the nuclear interactions, which not always fulfill the decoupling theorem.
It is undoutable that for a QCD-based precision nuclear physics approach, pions are an indispensable
ingredient, although in certain cases (few-nucleon reactions)  they can be integrated out and one still can make precise
calculations based on pionless EFT, see e.g. Refs.~\cite{Rupak:1999rk,Platter:2006ev,De-Leon:2022omx}, but 
this is more the exception than  the rule.

 Another playground of chiral dynamics is the investigation of fine-tunings in certain nuclear reactions,
 as the pion mass dependence can be used to set bounds on possible variations of the quark masses
 or the Higgs VEV. Here, input from lattice QCD at unphysical quark masses is very much needed,
 as Nature provides us only with one set of quark masses (and other fundamental constants).
 
 So can I answer the question posed in the title of this talk? The answer is yes and no. There are
 still a number of open ends, some of which I mentioned. Here, especially the lattice community
 can contribute significantly, however, the errors quoted  often appear too small in my view, which
 generates some (unnecessary) tensions. In particular, I would like to see more few-nucleon results
 at lower quark masses, which e.g. would serve to tighten the constraints on the variations of the fundamental
 parameters in nuclear reactions, besides from being interesting by themselves.  
 One could argue that chiral dynamics might become obsolete if lattice QCD operates at physical
 quark masses and has all the systematics under control. Looking at the topics discussed, I do not think
 that this will happen but rather one should continue combining the various theoretical tools to make
 the most precise predictions to be confronted by precision experiments.


\section*{Acknowledgements}
I thank all my collaborators who helped to shape my view on chiral dynamics
and the organizers for a great job done. The work reported  here is/was supported
in part supported by the  European Research Council (ERC) under the European Union's 
Horizon 2020 research and innovation programme (AdG EXOTIC, grant agreement No. 101018170),
the DFG through the Sino-German Collaborative Research Center CRC110 
``Symmetries and the Emergence of Structure in QCD'' (DFG Project ID No.~196253076 - TRR 110),
the CAS President's International Fellowship Initiative (PIFI) (Grant No.~2025PD0022) 
and the MKW NRW under the  funding code No.~NW21-024-A.

\end{document}